\documentstyle[prc,epsf,aps]{revtex}

\begin{document}

\title{Bridging Two Ways of Describing Final-State Interactions in
A($\mathbf{e,e'p}$) Reactions}

\author{D. Debruyne, J. Ryckebusch, S. Janssen and T. Van Cauteren}

\address{Department of Subatomic and Radiation Physics, \protect\\
Ghent University, Proeftuinstraat 86, B-9000 Gent, Belgium}

\date{\today}

\maketitle

\begin{abstract}
We outline a relativistic and unfactorized framework to treat the
final-state interactions in quasi-elastic A($e,e'p$) reactions for
four-momentum transfers Q$^{2}  \gtrsim 0.3$ (GeV/c)$^{2}$. The model,
which relies on the eikonal approximation, can be used in combination
with optical potentials, as well as with the Glauber
multiple-scattering method. We argue that such a model can bridge the
gap between a typical ``low'' and ``high-energy'' description of
final-state interactions, in a reasonably smooth fashion.  This
argument is made on the basis of calculated structure functions,
polarization observables and nuclear transparencies for the target
nuclei $^{12}$C and $^{16}$O.
\end{abstract}

\vspace{0.5cm}


\section{Introduction}

At intermediate values of the four-momentum transfer, here loosely
defined as Q$^{2} \geq 0.5$ (GeV/c)$^{2}$, the exclusive
electroinduced $\vec{e} + A \longrightarrow e' + (A-1)^* + \vec{p}$
reaction offers great opportunities to study the properties of bound
nucleons in a regime where one expects that both hadronic and partonic
degrees of freedom may play a role.  One such example is the study of
the short-range structure of nuclei.  These studies are meant to
provide insight into the origin of the large-momenta components in the
nucleus.  Amongst other things, constituent-quark models for the
nucleon predict measurable medium modifications of the bound nucleon's
properties.  At present, high-resolution double polarization
experiments of $A(\vec{e},e'\vec{p})$ reactions are putting these
predictions to stringent tests \cite{dieterich01}.  Another
medium-dependent effect, which has attracted a lot of attention in
recent years, is the color transparency (CT) phenomenon.  For
A($e,e'p$) processes, CT predicts that, at sufficiently high values of
Q$^{2}$, the struck proton may interact in an anomalously weak manner
with the spectator nucleons in the target nucleus \cite{frankfurt94}.

For all of the aforementioned physics issues, the interpretation of
the A($e,e'p$) measurements very much depends on the availability of
realistic models to describe the final-state interactions (FSI) which
the ejected proton is subject to.  There are basically two classes of
models to treat the FSI effects in electroinduced proton knockout. At
lower energies, most theoretical A$(e,e'p)$ investigations are
performed within the context of the so-called distorted-wave impulse
approximation (DWIA), where the scattering wavefunction of the struck
nucleon is calculated in a potential model \cite{picklesimer85}.  The
parameters in these optical potentials, which are available in both
relativistic and non-relativistic forms, are obtained from global fits
to elastic proton-nucleus scattering data.  The DWIA calculations
typically rely on partial-wave expansions of the exact solutions to
the scattering problem, a method which becomes increasingly cumbersome
at higher
energies. To make matters worse, global parametrizations of optical
potentials are usually not available for proton kinetic energies beyond
1~GeV.
In this energy regime the Glauber model \cite{glauber70}, which is a
multiple-scattering extension of the eikonal approximation, offers a
valid alternative for describing final-state interactions.
In such a framework, the effects of FSI are calculated directly
from the elementary proton-nucleon scattering data through the
introduction of a profile function
\cite{jeschonnek99,frankfurt95,benhar96,nikolaev95,ciofi99}.  Several
non-relativistic studies have formally investigated the applicability
of the Glauber model for describing A($e,e'p$) reactions at higher
energies and momentum transfers.  These investigations were often
hampered by the lack of high-quality A($e,e'p$) data to compare the model
calculations with. Recently, the first high-quality data for
$^{16}$O($e,e'p$) cross sections, separated structure functions and
polarization observables at Q$^2$ = 0.8~(GeV/c)$^2$ became available
\cite{gao00}.

The purpose of this letter is to investigate whether the optical
potential and the Glauber method for describing final-state interactions
lead to comparable results in an energy regime where both methods
appear applicable.  An observation which may point towards inconsistencies
in the description of FSI effects in A$(e,e'p)$ processes at ``low''
and ``high'' energies,
is the apparent Q$^{2}$ evolution of the extracted spectroscopic
factors \cite{lapikas00}.  Whereas numerous optical-potential analyses
of A($e,e'p$) measurements at low Q$^{2}$ have systematically produced
values which represent 50-70\% of the sum-rule strength,
it has recently been suggested that in order to describe the data at
higher Q$^{2}$ within the context of the Glauber model, substantially
higher values are required \cite{lapikas00,frankfurt00}.

We propose a relativistic formalism for computing A$(e,e'p)$
observables at medium energies.  The formalism is developed in such a
way that it can be used in combination with either optical potentials
or the Glauber method without affecting any other ingredient of the
model.  Results of optical potential and Glauber like calculations of
structure functions and polarization observables for the target nuclei
$^{16}$O and $^{12}$C are presented and compared.  In addition,
results of relativistic and unfactorized nuclear transparency
calculations for the $^{12}$C($e,e'p$) reaction are presented.

\section{Formalism}
\label{sec:formalism}

In the one-photon-exchange approximation, the cross section for a
process in which an electron impinges on a nucleus and induces the
knockout of a single nucleon with momentum $k_{f}$, leaving the
residual nucleus in a certain discrete state, can be written in the
following form \cite{raskin89}
\begin{eqnarray}
\left( \frac{d^{5}\sigma}{d\epsilon'd\Omega_{e'}d\Omega_{f}}
\right)
 =  \frac{M_pM_{A-1}k_{f}}{8\pi^{3}M_{A}} f_{rec}^{-1} \sigma_{M}
\biggl[ v_{L}{\cal R}_{L}+v_{T}{\cal R}_{T}+v_{TT}{\cal
R}_{TT}+v_{TL}{\cal R}_{TL} \biggr] \; ,
\end{eqnarray}
where $f_{rec}$ is the hadronic recoil factor, and $\sigma_{M}$ is the
Mott cross section. The electron kinematical factors $v_{i}$ and the
structure functions $R_{i}$ are defined in the usual manner
\cite{raskin89}.  Remark that in our model calculations an
unfactorized expression for the differential cross section is
adopted. This means that the off-shell electron-proton coupling is not
separated from the nuclear dynamics.  Although the factorized approach
has long been abandoned in the description of low-energy A($e,e'p$)
reactions, it is still widely used when it comes to describing
high-energy A($e,e'p$) processes.

In our model, the relativistic bound-state wavefunctions are
calculated within the context of a mean-field approximation to the $
\sigma - \omega$ model \cite{horowitz81,serot86}.  Assuming spherical
symmetry, the following type of solutions to the Dirac eigenvalue
problem result
\begin{equation}
\psi_{\alpha} ( \vec{r} ) \equiv \psi_{n \kappa m t} ( \vec{r} ) =
\left[
\begin{array}{c}
\imath G_{n \kappa t} ( r ) / r \; {\cal Y}_{\kappa m} \eta_{t} \\
- F_{n \kappa t} ( r ) / r \; {\cal Y}_{- \kappa m} \eta_{t}
\end{array}
\right] \; ,
\end{equation}
where $n$ denotes the principal, $\kappa$ and $m$ the generalized
angular momentum and $t$ the isospin quantum numbers. The ${\cal
Y}_{\pm \kappa m}$ are the well-known spin-spherical harmonics and
determine the angular and spin parts of the wavefunction.
In solving the relativistic bound-state problem, we have adopted the
values for the coupling constants and meson masses of Ref.~\cite{serot86}.

In the relativistic eikonal approximation, the scattering wave
function for a nucleon subject to a scalar ($V_s$) and a vector
potential ($V_v$) reads
\begin{eqnarray}
\psi_{\vec{k}_{f},s}^{(+)} = \sqrt{\frac{E+M}{2M}}
\left[
\begin{array}{c}
1 \\
\frac{1}{E+M+V_{s}-V_{v}} \vec{\sigma} \cdot \vec{p}
\end{array}
\right]
e^{\imath \vec{k}_{f} \cdot \vec{r}} e^{\imath S(\vec{r})}
\chi_{\frac{1}{2}m_{s}} \; ,
\label{eq:optical}
\end{eqnarray}
where the eikonal phase $S(\vec{b},z)$ is defined as
\begin{eqnarray}
\imath S(\vec{b},z) = - \imath \frac{M}{K} \int_{-\infty}^{z} dz' \,
\biggl[
V_{c} (\vec{b},z') + V_{so} (\vec{b},z') \left[ \vec{\sigma} \cdot
(\vec{b}
\times \vec{K} )- \imath Kz' \right] \biggr] ,
\end{eqnarray}
with $\vec{r} \equiv (\vec{b},z)$ and $\vec{K} \equiv \frac{1}{2}
(\vec{q}+\vec{k}_{f})$.  The central $V_c$ and spin-orbit potential
$V_{so}$ occurring in the above expression are determined by $V_s$ and
$V_v$ and their derivatives. In general, strength from the incident
beam is drained into other inelastic channels. Within the context of a
DWIA approach, the inelasticities are commonly implemented through the
use of a complex optical potential which is gauged against elastic
$pA$ scattering data.  In the numerical calculations, we have used the
global relativistic optical potentials of Cooper et
al.~\cite{cooper93}. By fitting proton elastic scattering data in the
energy range of 20 - 1040 MeV, Cooper \textit{et al.} obtained a set of
energy-dependent potentials for the target nuclei $^{12}$C, $^{16}$O,
$^{40}$Ca, $^{90}$Zr and $^{208}$Pb. In what follows, we refer to
calculations on the basis of Eq.~(\ref{eq:optical}) as the optical
model eikonal approximation (OMEA).

For proton kinetic energies $T_{p} \geq 1$ GeV, the use of optical
potentials appears no longer justifiable in view of the highly inelastic
character of the elementary proton-nucleon scattering process. Here, a
way out is offered by an extension of the eikonal method, namely the
Glauber multiple-scattering method, which is usually adopted in its
non-relativistic
version. Here, we propose the use of a relativized version
which allows us to write the wavefunction of the escaping proton as
\begin{eqnarray}
\psi_{\vec{k}_{f},s}^{(+)} = \sqrt{\frac{E+M}{2M}} \; \hat{\cal{S}}
\left[
\begin{array}{c}
1 \\
\frac{1}{E+M} \vec{\sigma} \cdot \vec{p}
\end{array}
\right]
e^{\imath \vec{k}_{f} \cdot \vec{r}}
\chi_{\frac{1}{2}m_{s}} \; .
\label{eq:glauber}
\end{eqnarray}
This expression for the relativistic scattering wave function is
derived in similar manner as for the non-relativistic (NR) case where the
wave function adopts the well-known form
\begin{eqnarray}
\psi_{\vec{k}_{f},s}^{(+),NR} = \hat{\cal{S}} \; 
e^{\imath \vec{k}_{f} \cdot \vec{r}} \chi_{\frac{1}{2}m_{s}} \; .
\end{eqnarray} 
The operator $\hat{\cal{S}}$ defines the action of the
subsequent collisions that the ejectile undergoes with the spectator
nucleons
\begin{eqnarray}
\hat{\cal{S}} (\vec{r},\vec{r}_{2},...,\vec{r}_{A}) = \prod_{j=2}^{A}
\left[ 1 - \Gamma (\vec{b} - \vec{b}_{j}) \theta(z-z_{j}) \right] \; ,
\end{eqnarray}
where $\theta(z-z_{j})$ ensures that the hit proton only
interacts with other nucleons if they are localised in its forward
propagation path.
The profile function $\Gamma(k_{f},\vec{b})$ for central elastic pN
scattering reads
\begin{eqnarray}
 \Gamma (k_{f},\vec{b}) = \frac{\sigma^{tot}_{pN}
(1-i\epsilon_{pN})}{4\pi\beta_{pN}^{2}} exp \left(
-\frac{b^{2}}{2\beta_{pN}^{2}} \right) \; .
\label{eq:profile}
\end{eqnarray}
The parameters in Eq.~(\ref{eq:profile}) can be taken directly from
nucleon-nucleon scattering measurements and include the total pN cross
sections $\sigma_{pN}^{tot}$, the slope parameters $\beta_{pN}$ and
the ratios of the real to imaginary part of the scattering amplitude
$\epsilon_{pN}$.  The A$(e,e'p)$ calculations on the basis of the
scattering state of Eq.~(\ref{eq:glauber}) are hereafter referred to
as the relativistic multiple-scattering Glauber approximation (RMSGA).

\section{Results}
\label{sec:results}

\begin{figure}
\begin{center}
{\epsfysize=10cm\epsffile{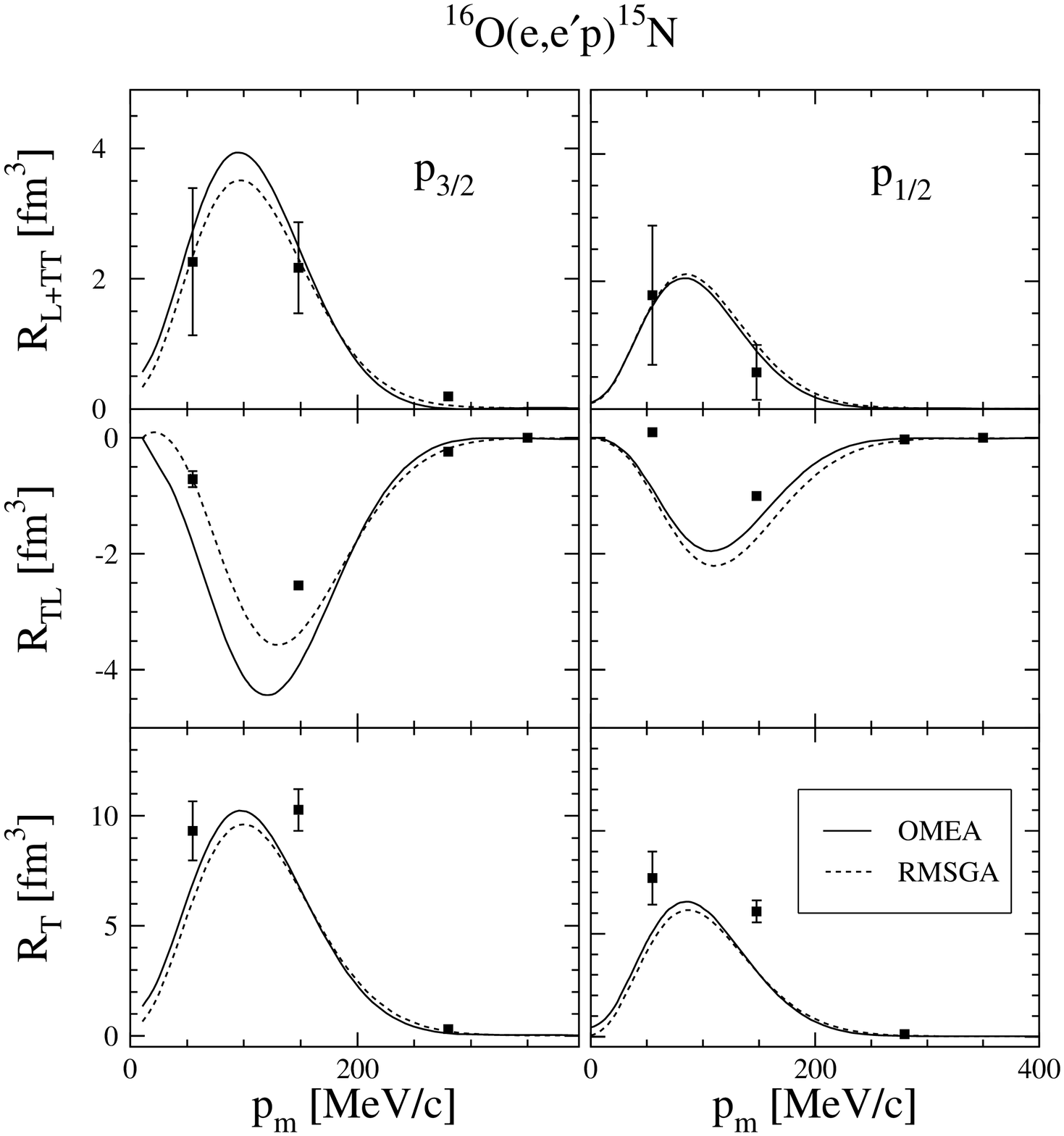}}
\end{center}
\caption{Separated structure functions for $^{16}$O$(e,e'p)$ in
quasi-perpendicular kinematics at $\epsilon$=2.4~GeV, $q$=1~GeV/c and
$\omega$=0.439~GeV. The solid and dashed curves denote the eikonal
(OMEA) and Glauber (RMSGA) calculations. The data are from
Ref.~\protect\cite{gao00}.}
\label{fig:struc}
\end{figure}

We first compare our relativistic calculations to recent quasi-elastic
$^{16}$O($e,e'p$) data from JLAB. In these high-resolution experiments
at Q$^{2} = 0.8$ (GeV/c)$^{2}$ differential cross sections, separated
structure functions and polarization observables were obtained
\cite{gao00}.  The variation in missing momentum was achieved by
varying the detection angle of the ejected proton with respect to the
direction of the momentum transfer (``quasi-perpendicular
kinematics''). Hence, it was only possible to isolate the combination
$R_{L+TT} \equiv R_{L} + \frac{v_{TT}}{v_{L}} R_{TT}$.  In
Fig.~\ref{fig:struc} we display the different structure functions
against the missing momentum $p_{m}$. The different curves all use the
same bound-state wavefunctions and electron-proton coupling but differ
in the way the FSI are treated.  In Table \ref{table:spec} we display
the spectroscopic factors which are obtained from a $\chi ^2$ fit from
the calculations to the data.  Although the Glauber and the
optical-potential framework provide an intrinsically very different
treatment of the final-state interactions, they lead to almost
identical spectroscopic factors at Q$^{2} = 0.8$
(GeV/c)$^{2}$. Another striking observation from Fig.~\ref{fig:struc}
is that both types of calculations produce almost identical results
for the $R_{T}$ and $R_{L+TT}$ structure functions.  In the $R_{TL}$
response for excitation of the 6.3~MeV $\frac{3}{2} ^-$ state the
differences are somewhat larger.  It is worth mentioning here that of
all structure functions, the $R_{TL}$ one has been identified as being
most sensitive to changes in the current operator and relativistic
corrections \cite{udias99,gardner94,jeschonnek98}.

\begin{figure}
\begin{center}
{\epsfysize=10cm\epsffile{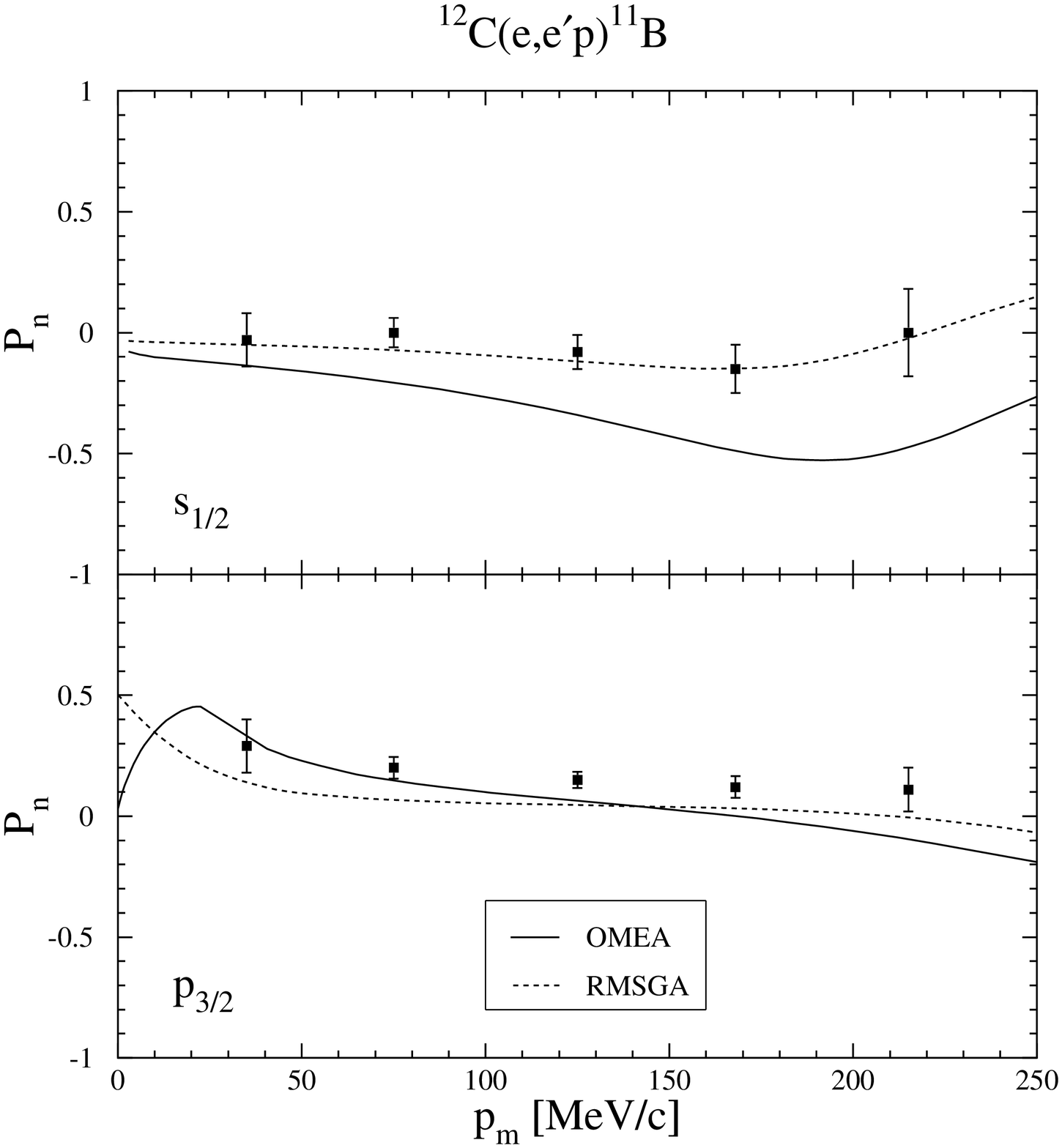}}
\end{center}
\caption{Induced polarization for the $^{12}$C($e,e'\vec{p}$)$^{11}$B
reaction in quasi-perpendicular kinematics at $\epsilon$=579~MeV,
$q$=756~MeV/c and $\omega$=294~MeV. The data are from
Ref.~\protect\cite{woo98}.}
\label{fig:pn}
\end{figure}

A quantity which is particularly sensitive to FSI effects is the induced
polarization $P_n$
\begin{eqnarray}
P_{n} = \frac{\sigma (s_{N}^{i} = \uparrow) - \sigma (s_{N}^{i} =
\downarrow)}{\sigma (s_{N}^{i} = \uparrow) + \sigma (s_{N}^{i} =
\downarrow)} \; ,
\end{eqnarray}
where $s_{N}^{i}$ denotes the spin orientation of the ejectile in the
 direction orthogonal to the reaction plane.  In the
 $^{12}$C($e,e'\vec{p}$) experiment of Woo and collaborators
 \cite{woo98}, the quantity $P_{n}$ was determined at quasifree
 kinematics for an energy and momentum transfer of ($\omega$,q) = (294
 MeV, 756 MeV/c). The results of the $^{12}$C($e,e'\vec{p}$)
 measurements are shown in Fig.~\ref{fig:pn}, along with our
 theoretical results.  The fair agreement of the Glauber results with
 the data is striking. It turns out that the applicability of the
 RMSGA method is wider than one would naively expect. We believe that
 the extended range of validity of the RMSGA method observed here, is
 (partly) caused by the relativistic and unfactorized treatment of the
 Glauber method. In Ref.~\cite{debruyne00} we have demonstrated that
 the effect of dynamical relativity (i.e.\ the effect of the lower
 components of the wavefunctions) can be significant at low Q$^{2}$.

\begin{figure}
\begin{center}
{\epsfysize=12cm\epsffile{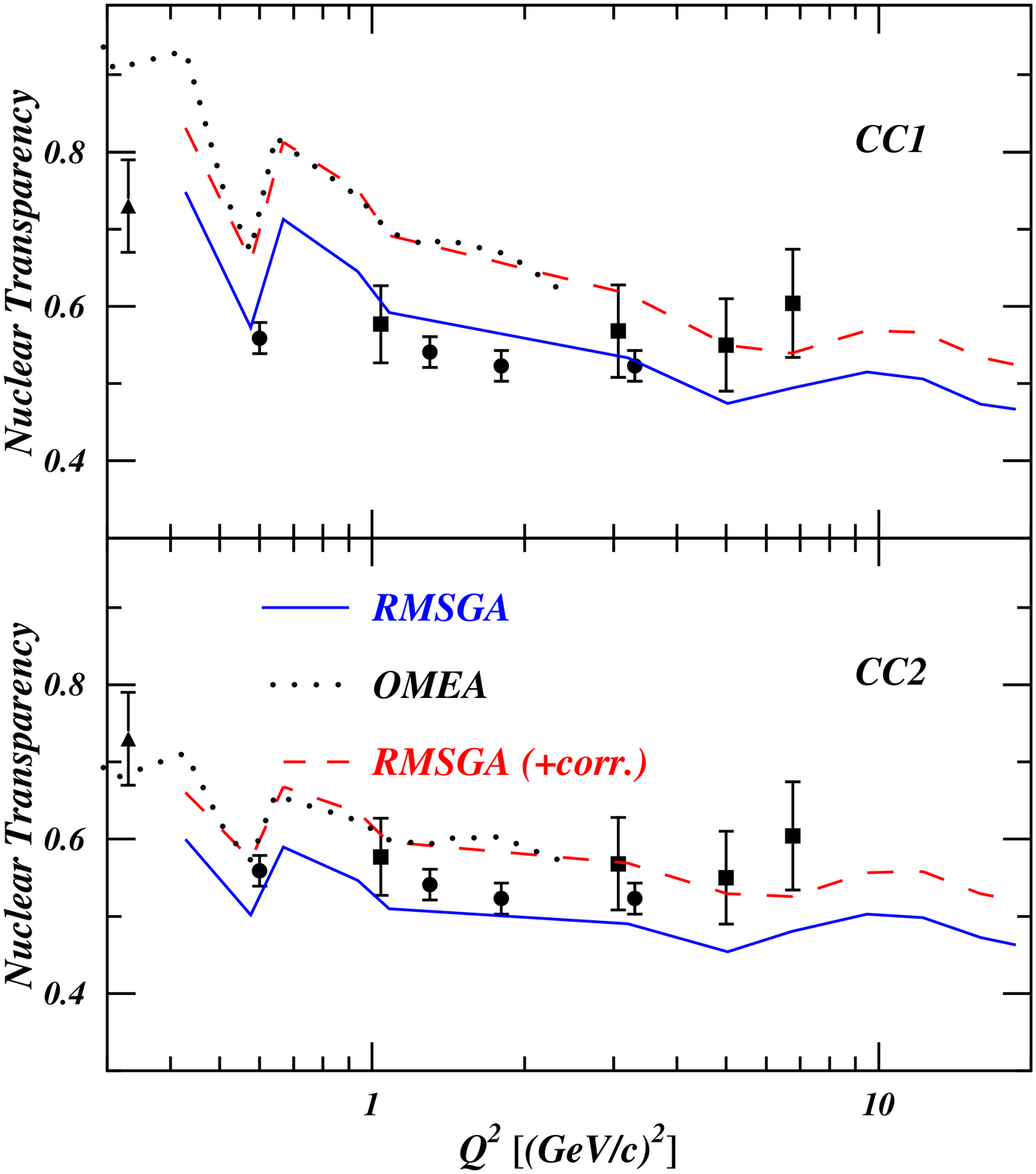}}
\end{center}
\caption{Nuclear transparency for $^{12}$C($e,e'p$) as a function of
Q$^{2}$. The curves denote calculations within the relativistic Glauber framework
(RMSGA) and the eikonal model with optical potentials (OMEA). Glauber results
with and without inclusion of SRC effects are presented. The
calculations in the upper (lower) panel employ the CC1 (CC2) current
operator. The curves assume full occupancy of the $s$ and $p$-shell
levels in $^{12}$C. The data are from BATES\protect\cite{bates}
(triangles), SLAC \protect\cite{makins94,oneill95} (squares) and from
JLAB \protect\cite{abbott98} (circles).}
\label{fig:trans}
\end{figure}

We now turn to the study of the nuclear transparency in the
quasi-elastic $^{12}$C($e,e'p$) reaction in a wide Q$^{2}$ range of
$0.3 \leq Q^{2} \leq 20$ (GeV/c)$^{2}$.  The results of our
calculations are contained in Fig.~\ref{fig:trans}.  We have performed
calculations within the relativistic Glauber framework and the eikonal model with
the optical potentials from Ref.~\cite{cooper93}.  The optical potential
results are limited to kinetic energies below $T_{p} = 1$ GeV.  In the
Glauber model, we have also performed calculations which include the
effect of short-range correlations (SRC).  Each of these calculations
was done with the CC1 and CC2 current operator.  The transparencies
calculated within the Glauber framework exhibit some fluctuations.  The
magnitude of these fluctuations mark the intrinsic uncertainties on
the computed transparencies caused by the error bars on the measured
elementary pN scattering parameters.  Most published Glauber
calculations for the nuclear transparencies do not exhibit these
fluctuations, but use (smooth) global fits to determine the energy
dependence of the elementary pN scattering data.  In our numerical
calculations, we use the listed experimental pN results.

The measurements in Refs.~\cite{makins94,oneill95,abbott98} were
performed in certain regions of the phase space, dictated by the
requirement that quasi-elastic conditions should be met. We
have constrained our calculations to the same  segment of the phase
space. In general, the experimental transparency $T_{exp}$ is defined as
\begin{eqnarray}
T_{exp} = \frac{\int_{\Delta^{3}k} d\vec{k} \int_{\Delta E} dE \;
\left(
\frac{d^{5}\sigma}{d\epsilon'd\Omega_{e'}d\Omega_{f}}
\right) _{exp}}
{ c_{A} \; \int_{\Delta^{3} k} \int_{\Delta E} dE
\;
\left(
\frac{d^{5}\sigma}{d\epsilon'd\Omega_{e'}d\Omega_{f}}
\right) _{PWIA}
}
 \; .
\label{eq:formuletrans}
\end{eqnarray}
The A-dependent factor $c_{A}$ renormalizes the non-relativistic
plane-wave impulse approximation (PWIA) predictions for
corrections induced by SRC.  For the $^{12}$C($e,e'p$) process, a
correction factor
of $0.901 \pm 0.024$ was adopted in
Refs.~\cite{makins94,oneill95,abbott98}.
As the implementation of short-range correlations can be done in
numerous ways, we have removed this factor and rescaled the data
accordingly.

At lower values of Q$^{2}$ there are substantial deviations between
the transparencies computed with the CC1 and the CC2 current
operator. At higher values of Q$^{2}$ [Q$^{2} \geq 3 - 4$
(GeV/c)$^{2}$], where the differences between the CC1 and CC2
predictions are negligible, the predicted transparencies tend to
underestimate the measured transparencies, even when assuming full
occupancy for the single-particle levels in $^{12}$C.  This apparent
shortcoming can be cured by introducing the effect of short-range
correlations.  They are implemented through the introduction of a
central (or, Jastrow) correlation function $g(\mid \vec{r_1} -
\vec{r_2} \mid)$ in the two-body density components which are part of
the Glauber calculations.  In practice, this procedure amounts to the
following replacement in the matrix elements that determine the
rescattering effects
\begin{equation}
\psi_{\alpha} ( \; \vec{r_1} \;  ) \psi_{\beta} ( \; \vec{r_2} \; )
\longrightarrow 
\psi_{\alpha} ( \; \vec{r_1} \;  ) \psi_{\beta} ( \; \vec{r_2} \; )
g(\mid   \vec{r_1} - \vec{r_2} \mid) \; .
\end{equation}

In line with the findings of other studies
\cite{frankfurt95,frankfurt00,kohama93,benhar92,pandharipande92} we
observe that short-range correlations increase the calculated
transparencies by about 10 \% .  We have used the central correlation
function $g_{GD}(r)$ from a nuclear-matter calculation of Gearhart and
Dickhoff \cite{gearheart94}.  Amongst different other candidates, this
correlation function emerged as the preferred choice in an analysis of
$^{12}$C($e,e'pp$) data \cite{blomqvist98}.  Being a rather hard
correlation function, the effect of introducing $g_{GD}(r)$ on the
computed values of the transparencies is maximized.  We have
also evaluated the role of relativistic effects on the computed
transparencies.  In general, these effects are rather small.  For
example, the coupling between the lower component in the bound and
scattering state marginally affects the predictions for the
transparencies.  For some specific observables, like the $R_{TL}$
structure function in Fig.~\ref{fig:struc}, on the other hand, the
relativistic effects are substantial.

For four-momentum transfers about Q$^{2} \approx 1$ (GeV/c)$^2$ it
appears legitimate to directly compare the optical potential and the
Glauber calculations.  It is obvious from Fig.~\ref{fig:trans} that
the OMEA curves exhibit a behaviour very similar to the correlated
Glauber results.  This observation may suggest that the in-medium pN
cross sections are modestly reduced compared to the on-shell values
and that the major part of this effect can be modeled through the
introduction of SRC mechanisms.  As for the OMEA results, it can be
argued that the SRC effects, which belong to the class of medium effects,
are already effectively incorporated in the formalism.  After all, the
optical potentials are obtained from a global fit to proton-nucleus
scattering data.

The apparent consistency between the OMEA and the correlated RMSGA
predictions is an interesting result.  Indeed, it demonstrates that
the low and the high Q$^{2}$ regime can be bridged in a satisfactory
manner.  This feature has it consequences for the apparent Q$^{2}$
evolution of the spectroscopic factors extracted from
$^{12}$C$(e,e'p)$ \cite{lapikas00}.  As suggested by the authors of
Ref.~\cite{lapikas00}, a consistent analysis of all $^{12}$C($e,e'p$)
data between $0.1 \leq Q^{2} \leq 10$ (GeV/c)$^{12}$ could much
improve insight into this matter. A consistent treatment would at
least allow to separate genuine physical effects (contributions from
meson exchange, $\Delta$-isobars, SRC etc.) from model-dependent
uncertainties (like gauge ambiguities and problems related to the
treatment of the FSI).  Such an analysis should preferably be carried
out in a framework that is able to describe both the low and high
Q$^{2}$ $(e,e'p)$ data without any inconsistencies in some
intermediate-energy range. We feel that the framework presented here,
is an initial step in this direction.

\section{conclusion}
\label{sec:conclusion}

Summarizing, we have outlined a relativistic and unfactorized
framework for computing A($e,e'p$) observables at intermediate and
high four-momentum transfers Q$^{2}$.  The model is based on the
eikonal approximation and can accomodate both relativistic optical
potentials and a Glauber approach, which are two substantially
different techniques to deal with final-state interactions.  We have
shown that optical-potential and Glauber predictions are reasonably
consistent at intermediate values of Q$^{2}$.  Indeed, at Q$^{2}$
values of about 0.8 (GeV/c)$^{2}$, which is a regime in which both
approaches for dealing with the FSI's appear justified, comparable
results for the transparencies and structure functions are obtained.

\begin{table}
\begin{center}
\begin{tabular}{l|lll|lll}
& \multicolumn{3}{c}{\bf{CC1 operator}} & \multicolumn{3}{c}{\bf{CC2
operator}} \\ \hline
& \em{RPWIA} & \em{OMEA} & \em{RMSGA} & \em{RPWIA} & \em{OMEA} &
\em{RMSGA} \\ \hline
$\frac{3}{2} ^-$ ($E_x$=6.3~MeV) & 0.59 & 0.96 & 0.96 & 0.61 & 1.00 & 1.00
\\ \hline
$\frac{1}{2} ^-$ (g.s.) & 0.53 & 0.79 & 0.80 & 0.53 & 0.82 & 0.82
\end{tabular}
\end{center}
\caption{The spectroscopic factors as derived from the
$^{16}$O($e,e'p$) results contained in Fig.~\protect{\ref{fig:struc}}
through a $\chi^{2}$ fitting procedure.}
\label{table:spec}
\end{table}

\end{document}